\documentclass[prodmode,acmcsur]{acmsmall} 

\begin{document}

\markboth{M. Tsvetkova et al.}{Understanding Human-Machine Networks}

\title{Understanding Human-Machine Networks: A Cross-Disciplinary Survey}
\author{MILENA TSVETKOVA
\affil{Oxford Internet Institute, University of Oxford}
TAHA YASSERI
\affil{Oxford Internet Institute, University of Oxford}
ERIC T.~MEYER
\affil{Oxford Internet Institute, University of Oxford}
J.~BRIAN PICKERING
\affil{IT Innovation Center, University of Southampton}
VEGARD ENGEN
\affil{IT Innovation Center, University of Southampton}
PAUL WALLAND
\affil{IT Innovation Center, University of Southampton}
MARIKA L\"{U}DERS
\affil{SINTEF}
ASBJ{\O}RN F{\O}LSTAD
\affil{SINTEF}
GEORGE BRAVOS
\affil{Athens Technology Center}
}

\begin{abstract}
In the current hyper-connected era, modern Information and Communication Technology
systems form sophisticated networks where not only do people 
interact with other people, but also machines take an increasingly visible and participatory 
role. Such human-machine networks (HMNs) are embedded in the daily lives of people, both for 
personal and professional use. They can have a significant impact by producing synergy and innovations. 
The challenge in designing successful HMNs is that they 
cannot be developed and implemented in the same manner as networks of machines nodes alone, nor 
following a wholly human-centric view of the network. The problem requires an interdisciplinary approach.
Here, we review current research of relevance to HMNs across many disciplines. 
Extending the previous theoretical concepts of socio-technical systems, actor-network theory, 
cyber-physical-social systems, and social machines, 
we concentrate on the interactions among humans and between humans and machines. We identify 
eight types of HMNs: public-resource computing, crowdsourcing, web search engines, crowdsensing, 
online markets, social media, multiplayer online games and virtual worlds, and mass collaboration. 
We systematically select literature on each of these types and review it with a focus 
on implications for designing HMNs. Moreover, we discuss risks associated with HMNs 
and identify emerging design and development trends.
\end{abstract}

%
%


%
%



\begin{bottomstuff}
This project has received funding from the European Union's Horizon 2020 research and innovation program under grant 
agreement No 645043.

Author's addresses: M. Tsvetkova, T. Yasseri, and E.T. Meyer, Oxford Internet Institute, University of Oxford, Oxford OX1 3JS, UK; J.B. Pickering, V. Engen, and P. Walland, IT Innovation Center, University of Southampton, Southampton SO16 7NP, UK; 
M. L\"{u}ders and A. F{\o}lstad, SINTEF, 0373 Oslo, Norway; G. Bravos, Athens Technology Center, Chalandri 152 33, Greece.
\end{bottomstuff}

\maketitle

\section{Introduction}

Since the invention of the Gutenberg press, machines have had a transformative impact on human interaction. 
As a result of the Industrial Revolution, machine influence on how we communicate, exchange, and cooperate with each other 
has become more immediate and pervasive. Today, with the ever increasing proliferation of networked technology 
in support of human and machine interaction, human actions and interactions have become so interrelated with 
technology that it is difficult to know whether society changes because of technology or the other way around.

Increasingly, the interactions of humans and machines form interdependent networks. We conceptualize these 
networks as human-machine networks (HMNs), that is, assemblages of humans and machines that interact 
to produce synergistic effects. Ever more activities in work and private life are conducted within 
HMNs. For example, actions to address environmental problems are executed in networks 
involving government, interest organizations, citizens, smart devices, and sensor networks. Systems for emergency 
response and rescue involve complex interactions between sensors, smart machines, victims, volunteers, and 
emergency response organizations. Scientific research is increasingly conducted in networks of scientists, 
crowdsourced volunteers and amateurs, and networked machine resources, and many more examples.

In consequence, HMNs increasingly influence our society. For the individual worker and citizen, 
the form, experience, and outcome of life may sometimes depend less on the characteristics of the individual and more 
on the characteristics of the online and offline networks in which they are embedded. 
For companies, the public sector, and organizations, 
productivity, innovativeness, and civic participation depend on the characteristics of the networks of which 
workers and citizens are part. Hence, an advanced understanding of HMNs, and how to benefit 
from their characteristics, is important to strengthen productivity and innovation, as well as citizen 
participation and quality of life.

The challenge is that HMNs cannot be developed and implemented in the same manner as networks 
of machine nodes alone. Creating successful solutions for HMNs requires awareness concerning 
the kind of network to be established, the actors that are involved, the interactions and the 
capabilities and behaviors emerging within the network. By establishing such awareness, we may benefit
from the existing knowledge and experience when designing the HMNs of the future.

There are four major fields of research of relevance to our understanding of HMNs: 
Socio-Technical Systems (STS) theory, Actor-Network Theory (ANT), Cyber-Physical-Social Systems (CPSS) theory, and the emerging theory on social machines. STS
theory provides insight into the dual shaping of technology and the social 
(work) context in which it is implemented \cite{Leonardi2012}, recognizing organizations as complex systems of humans 
and technology that aim to reach given goals in the context of a given organizational environment \cite{Emery1965}. 
ANT aims at fundamentally revising what should be considered part of the social, and argues 
that we explicitly need to take into account that any social system is an association of heterogeneous elements 
such as humans, norms, texts, devices, machines, and technology \cite{Latour2005,Law1992}, granting equal weight 
to humans and non-human (machine) entities in the analysis of the social. Hence, whereas STS theory considers 
the social as part of technology, ANT considers technology (and other non-human entities) as part of the social. 
More recently, CPSS theory has extended the STS approach and called attention to the human and social dimensions present in systems that use computational algorithms to control or monitor physical devices \cite{Wang2010,Sheth2013}.
Representing a fourth perspective, the emerging theory on social machines refers to 
systems that combine social participation with machine-based 
computation \cite{Buregio2013,Shadbolt2013,Smart2014}, connecting with Berners-Lee's original vision of 
the Web more as a social creation than a technical one, ``to support and improve our Web-like existence 
in the world'' \cite[p. 123]{Berners-Lee2000}.

However, even if these theories conceptualize humans and machines as forming a single system, as opposed to perspectives where social structures are seen as merely mediated in  machine networks (e.g. social network theory), they do not provide the insight and guidance needed to support the design of contemporary and future HMNs. Furthermore, 
these theories do not contribute to a unified framework for understanding HMNs, as is seen from the plethora of theoretical positions in which current studies of such networks are based. As will be seen in Section~\ref{sec:networks}, the volume of publications concerning HMNs is rapidly increasing across a wide range of academic fields. Such breadth in academic attention is beneficial for the development of knowledge on HMNs. However, this breadth also implies a potential risk for fragmentation, barring opportunities for the transfer of knowledge and experiences across academic, as well as practical fields of study.       

To initiate the broad theoretical basis needed to bridge the rapidly evolving field of HMNs, 
we have conducted a cross-disciplinary literature survey 
of key studies that concern different types of HMNs. 
Multiple surveys of particular types of HMNs exist \cite{Crowston2012,Guo2013,Yahyavi2013,Guo2015,Pejovic2015} 
but we are the first to present a comprehensive overview of the field within a unifying framework.  
The objective of our paper 
is to systematically review the existing literature, with a specific focus on the human and machine actors 
and the interactions between them. This analysis provides a first step towards integrating current academic efforts to understand the emerging phenomenon of HMNs. Furthermore, the analysis demonstrates how the identification of key characteristics in different HMNs may serve to expose issues related to design of HMNs, as well as to support the transfer of knowledge and experience across academic disciplines and types of HMNs.

In Section~\ref{sec:scope-approach} we discuss the scope and approach to the systematic literature review we have conducted. In Section~\ref{sec:networks} we present real-world examples and insights from the literature on different types of human machine networks. We discuss common challenges, emergent behavior, and trends in Section~\ref{sec:discussion}. We conclude in Section~\ref{sec:conclusions}.


\section{Background and scope}
\label{sec:scope-approach}

In this section, we provide a definition of an HMN, discuss background literature on the actors that make up HMNs and their interactions, and establish the scope and approach taken to conduct this survey. 

\subsection{Definition}

HMNs are assemblages of humans and machines whose interactions have synergistic effects. This means that the effects generated by the HMN should exceed the combined effort of the individual actors involved. In particular, the HMN should serve a purpose which would not have been achieved merely by the effects of the individual actors. Typically, the synergistic effect is achieved by combining human strengths and machine strengths. Interactions in the HMNs 
are expected to result in outputs that neither a pure social network, nor a computer network could achieve independently. 

The concept of HMNs is to be understood as a perspective for analysis or conceptualization. Given our concern for synergistic effects, the networks of main interest are those where the synergistic effects between humans and machines are 
immediately evident such as in systems for mass collaboration (see Section~\ref{sec:masscollaboration}). Networks of lesser interest include, for example, 
simple communication and broadcasting networks such as telephone, telegraph, e-mail, or TV networks, as the medium here 
may be said to hold more the role of a non-interfering intermediary. In other words, for a network to be considered 
an HMN, the machines need to transform and/or influence, not just transmit. In the parlance of actor-network theory, we focus on 
networks where machines serve as mediators, by affecting and transforming the interaction that takes place. 

\subsection{Actors}

The two types of actors in HMNs are the human and the machine. By ``human'' we mean an entity that 
behaves like a single person (even if the entity is an organization). In contrast to machines, humans have the capacity 
for emotions, attitudes, sociality, meaning-making, creativity, and intent. Compared with machines, human behavior is typically more unpredictable, yet susceptible to influence.

In most cases, any person can participate in an HMN. Participation is more often determined by the benefits and costs the individual perceives than the 
individual's social status or geographic location. The benefits can be divided into 
economic, intrinsic, and social \cite{Ardichvili2008,Chiu2006,Yee2006}. Individuals profit economically when they 
receive payments (in money, goods, or services) or obtain skills and qualifications for their paid career. They gain 
intrinsically when they contribute to a good cause, share knowledge, or help others. 
Some are also intrinsically motivated to gain or create knowledge for its own sake. Individuals 
may also benefit socially if participation in the HMN allows them to create, build, and maintain reputation and social 
capital, or simply enjoy social interaction. In general, people often have complex motivations to participate and 
contribute and motivation strategies such as competition and gamification make use of this by offering a combination of economic, intrinsic, and social benefits. Time and effort are the two obvious costs that people incur when they participate in HMNs. Another significant 
cost is the risk of breached privacy \cite{Krasnova2010,Sheehan2002}.  Leaking private information to the public and 
stealing a user's identity can entail significant financial losses, family distress, and social embarrassment. 

Regarding machines, any device with connectivity can be part of an HMN:  personal computers, smartphones, 
tablets, wearable technology, sensors, embedded chips, servers running algorithms, and so on. These machines take input 
in the form of text, media, or sensor data, such as vital signs or environmental measures, printed or coded 
instructions and software, signals and alerts from multiple sources, including other machines as well as humans. This 
input can be entered by the user or collected automatically. A machine with computational capabilities can aggregate, 
clean, or otherwise transform the input data in order to output something else or even fuse data from different sources 
to make complex decisions to be interpreted by other machines or by human experts \cite{Atzori2010,Castells2011,Lee2015}.

To facilitate the functioning of the HMN, the machines need to be available, 
connected, and secure. Compared to humans, the machines do not have motivation, do not experience 
trust or reliance, and do not behave altruistically or irrationally of their own volition: they have no agency 
in that sense \cite{Jia2012,Rose2005}. 
With autonomic computing, however, there may be a promise of some level of self-regulation and `healing' 
\cite{Huebscher2008,Kephart2003,White2004}. Moreover, machines are increasingly capable of solving complex 
data analysis problems which humans would struggle to tackle not least at such speed, although affective responses 
and the subtlety of nuanced behaviors remain elusive \cite{Norman2003,Pantic2003,Picard2001}. Further, machine learning and data mining crucially depend on large data sets generated and annotated by humans. In such cases, a 
quasi-balanced collaboration is required. Machines are therefore now becoming increasingly significant contributors 
to HMNs, and no longer simply communication channels or facilitators of highly-distributed human-to-human connectivity.

\subsection{Interactions}

Machines enable new forms of interaction between humans. Different HMNs allow participants to interact with different 
levels of intensity and involvement. In some networks (for example, crowdsourcing platforms such as Amazon Mechanical 
Turk), participants do not have the opportunity to directly interact. In others, they can 
observe each other's contributions. Depending on the application, they can browse all contributions (as in an online 
market such as eBay), see algorithmically selected contributions (as in 
Reddit, which is a content site), see algorithmically customized contributions (for example, 
via the Newsfeed algorithm on Facebook), or only view aggregate results (as in prediction 
markets). Once being able to observe contributions, participants can then evaluate them by approving them (for example, 
``likes'' on Facebook), approving or disapproving them (``upvotes'' and ``downvotes'' on Reddit), rating them 
(Amazon product ``stars''), or commenting on them. Finally, participants can sometimes modify 
others' contributions by editing or deleting them. For example, Wikipedia editors can revert edits and Linux developers 
can fix code contributed by others. Human-human interactions largely depend on trust \cite{Dwyer2007,Jones2008}. 
Phenomena such as trolling, cyber-bullying, and cyber-stalking signify the erosion of trust. Human-human interactions 
also involve social influence \cite{Bond2012,Muchnik2013,Onnela2010}. Social influence can lead to large-scale 
behavioral or emotional contagion. This can be both positive, if it results in the spread of pro-social or 
health-conscious behavior, and negative, if it leads to dangerous herd behavior or unproductive groupthink.

When the role of the machine in an HMN may be seen as that of a mediator (that is, its output does not equal the input 
from the other nodes in the HMN), it is relevant to analyze the human-machine interaction. To some degree, human-machine 
interactions are gated only by the communication protocol used. Input can be provided via different channels 
including keyboards, data channels, speech, and movement and gesture, actively or even passively. Interactions in HMNs are therefore 
increasingly multi-modal, with different HMNs allowing individuals to contribute or interact in different ways 
sometimes dictated only by ease of use \cite{Lin2003,Lin2003a}. Similarly, the output channel may 
depend on the service or application (signage, broadcasting, experimentation, web searches, and so forth) or on the preferences of the recipient (a recorded message, a printout, or a visual display).

In respect of what humans input, individuals make contributions to different extents. In some networks, users do 
not have to do anything to contribute data beyond interacting with the service or even simply turning on the service. In 
these cases, participants contribute passively. Their social interactions, personal characteristics, and site 
behavior are automatically collected by the service. These data are then analyzed and often used to recommend actions 
(as some crowdsensing applications do), social contacts (as in LinkedIn), or products (as 
in Netflix). In other HMNs, contributors need to make the effort and take the time to 
contribute actively. They contribute by filling out surveys, composing and editing text, writing code, sending sensor 
measurements, or uploading videos, photos, and news links. In more extreme cases, participants can also modify the 
content on the HMN or the rules by which the content is collected, managed, and distributed. Consumers then 
become ``prosumers,'' and can still switch between the two modes as suits the service or their own needs. 

In respect of what machines output, different HMNs may introduce different levels of control over the content. 
Machines may simply list or show the contributions, as is common in online marketplaces and 
in direct response to a specific query. Alternatively, they may employ complex algorithms to filter 
and select contributions, as Web search engines do, for example, personalising or customising results 
to reflect previous user behaviors. Or they may even use bots and human intelligence to modify 
contributions. For example, Wikipedia uses vandalism-detection algorithms and forums employ censorship 
algorithms, while content sites like Digg maintain centralized content management. 

\subsection{Scope and approach}

To narrow the scope of this survey, we simplify the analysis of machine-machine interactions. In 
particular, when the underlying or enabling machine networks are perceived as one and the same unit by all 
the relevant nodes of the HMN, we analyze and represent the machine sub-network as a single node. This is not to 
say that machine-machine interactions will not be considered; rather, they 
will be considered only to the extent that is valuable to the analysis at hand.

Thus, our discussion focuses on the human-human and human-machine interactions in the context of contemporary HMNs. 
There are clearly differences in the relative roles and status of machines and human actors 
across the different types of network. For instance, there will be cases where the machines act as a ``human proxy'' to 
facilitate what is essentially human-to-human relationships. In other cases, they become semi-autonomous agents that 
act alongside and in collaboration with human agents in the network. We base our analysis on the 
existence and intensity of these interactions and the extent to which they are essential for the functioning of the 
HMN.

\begin{figure}
\centerline{\includegraphics[scale=0.65]{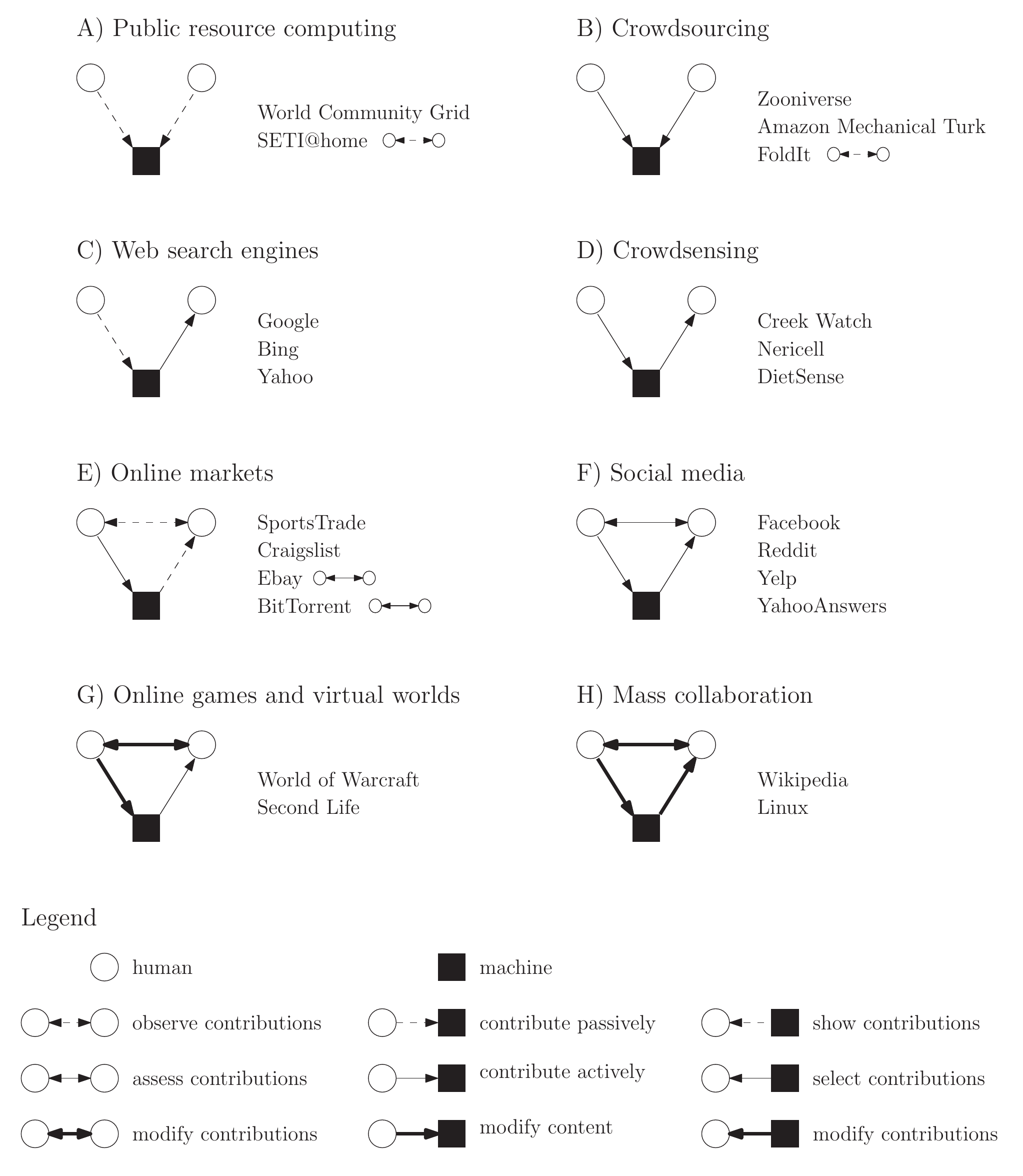}}
\caption{Types of human-machine networks.}
\label{fig:types}
\end{figure}

Based on different possible permutations of interaction types, we identify and differentiate between eight types of HMNs: public resource computing, crowdsourcing, Web search engines, crowdsensing, online markets, social media, multiplayer 
online games and virtual worlds, and mass collaboration (Figure~\ref{fig:types}). These types are to be seen as an 
organizing framework for the survey, not as a complete taxonomy of HMNs. We used keywords associated
with these types to systematically select relevant literature, with a focus on recent and high-impact articles.
Details of our selection procedure are described in the online appendix. 

\section{Analysis of human-machine networks}
\label{sec:networks}

\begin{figure}
\centering
\includegraphics[scale=1]{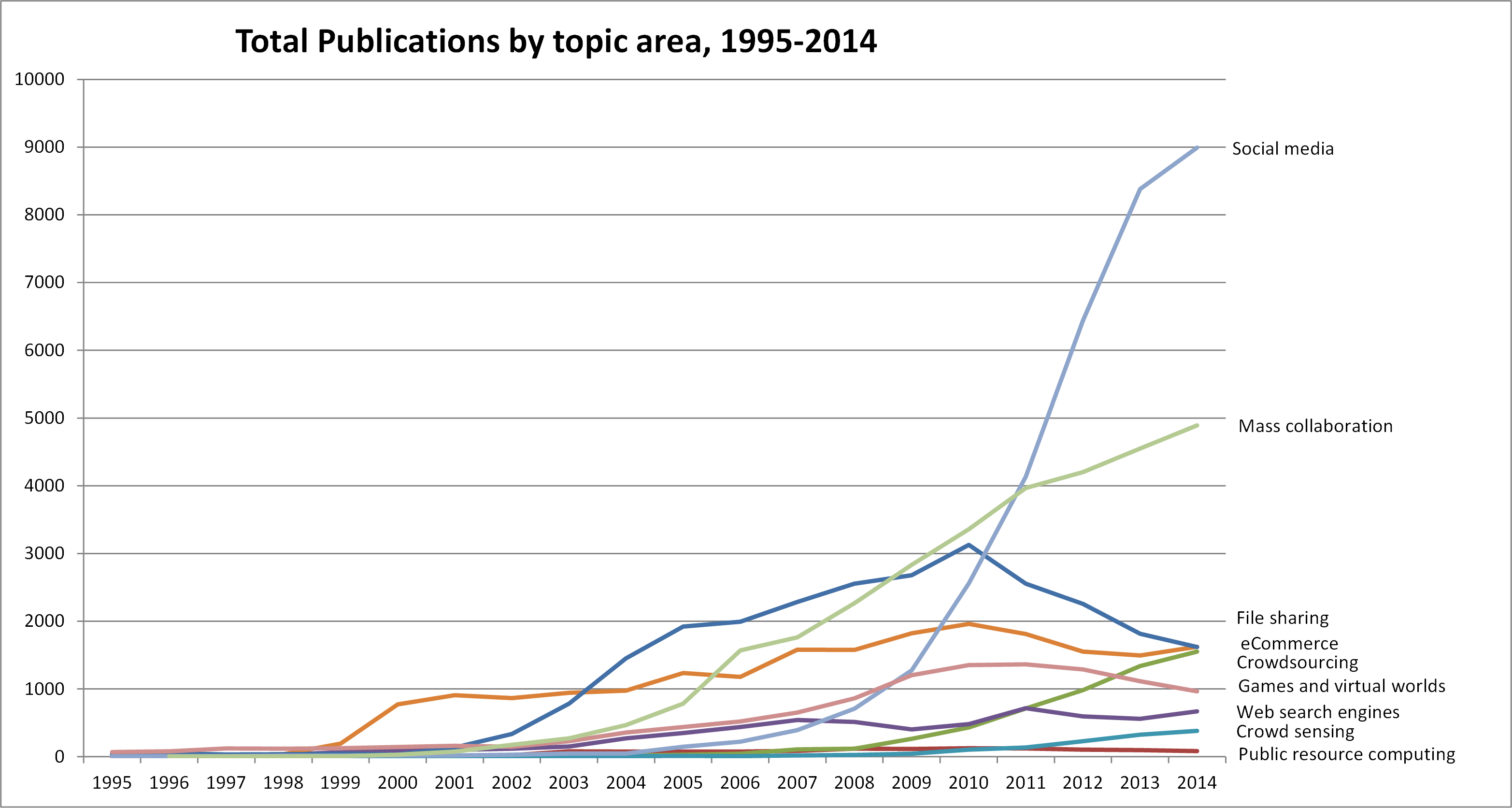}
\caption{Publications related to human-machine networks.}
\label{fig:pubs}
\end{figure}

The number of publications on HMNs, as measured by the number of articles retrieved by our keywords from the Scopus bibliographical database (www.scopus.com), has increased over the last 20 years (Figure~\ref{fig:pubs}). More importantly, the pattern of publications has changed over time. During the late 1990s, relatively few publications used the particular terms in their title, keywords, or abstract. The earliest increase was in the eCommerce area (part of the online markets type), where publications increased from 192 in 1999 to 772 in 2000, and remained near or above 1000 publications per year after that. The second area to gain prominence is file sharing, which increased from less than 50 publications annually in 2000 to approximately 2000 per year by 2005.

The pattern for social media is particularly striking in the chart, showing a steep rise starting in 2008 (n=708) and accelerating by 2010 (n=2560) to a high of 8991 publications by 2014. Mass collaboration as a topic, on the other hand, shows a more steady rise over the last ten years, peaking at 5165 publications in 2014.

\begin{figure}
\centering
\includegraphics[scale=1]{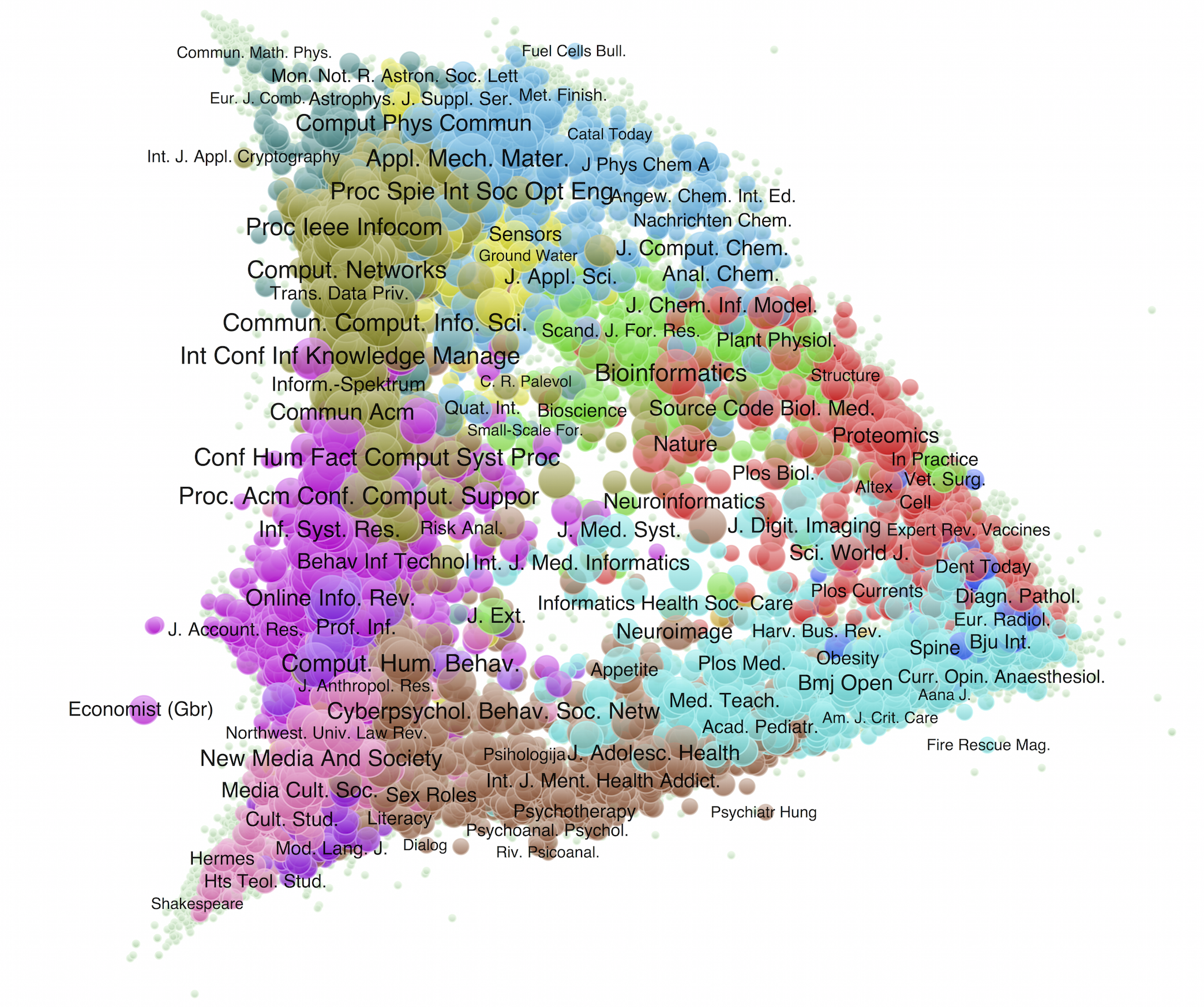}
\caption{Journals publishing articles related to human-machine networks. The colored dots are Scopus journals that include articles with the key terms we identified; the background uncolored dots do not include such articles. The colors represent broad ``community'' groupings distinguished by the visualization software. Larger dots with corresponding larger labels have more published articles, and smaller dots/labels have fewer.}
\label{fig:journaloverlay}
\end{figure}

We next look at the extent to which ideas relating to human-machine interaction have penetrated various disciplines as indicated by the journals where these publications appear. In Figure~\ref{fig:journaloverlay}, we use the methods described in \citeN{Leydesdorff2013} and \citeN{Leydesdorff2014} to create an overlay map of the journals in the data extracted from Scopus. The method involves extracting the journal names and processing them with Leydesdorff's tools. \footnote{The tools are available at \url{http://www.leydesdorff.net/scopus_ovl/}.} These tools match our Scopus data with a standardized map of science generated using journal-journal citations to discover the closeness of journals to each other, and thus the relative location of disciplines. The underlying map thus represents ``all of science,'' or at least as much of it as is represented in the 19,600 journals indexed by Scopus.

We observe that HMNs, as defined by our keywords, spread across all areas of science. There is a concentration one might expect in the upper left portion of the figure in the computer science-related journals, but there is nearly as much publication activity in the physical sciences at the upper right, the medical sciences at the lower right, and in the social sciences and (to a lesser extent) the humanities in the lower left.

In what follows, we present the most prominent articles from this interdisciplinary literature, 
related to human participation and to human-human and human-machine interactions. 

\subsection{Public resource computing}

Public resource computing (PRC) projects are distributed computing networks, where each processing unit is a 
personal computer voluntarily embedded in the network to collaboratively solve computational problems. 
The majority of PRC projects are dedicated to scientific projects in fields as 
diverse as astrophysics, mathematics, molecular biology, and seismology, where extremely computation-intensive tasks 
are common. PRC networks are the simplest HMNs as far as the human-human and human-machine interactions are concerned
(Figure~\ref{fig:types}A). Once volunteers sign up for a PRC project, they are no 
longer required to actively contribute time or effort apart from what their CPU is contributing. At the same time, 
they will not receive any substantial feedback from the project either.
The first PRC systems were launched in 1996-1997: GIMPS (The Great Internet Mersenne Prime Search) and Distributed.net 
\cite{Anderson2002}. SETI@home is another early instance of a PRC network. It was launched in 1999 with the goal to 
analyze radio signals from space in search of extra-terrestrial intelligent life. Further examples include the World 
Community Grid, intended to tackle scientific research projects in any field \cite{Hachmann2011} and GPUGRID, 
designated to molecular dynamics simulation methods \cite{Buch2010}.

The design of PRC systems crucially depends on building the machine-machine interaction infrastructure. Nevertheless, 
considerations regarding the recruitment and motivation of contributors remain important. To attract participants, 
successful PRC platforms have emphasized the social value of their scientific projects and let mass media news coverage 
and word-of-mouth do the rest \cite{Anderson2002}. To retain participants and increase contributions, they have built 
reputation systems based on numeric measures of contributed computation and allowed the ability to view others' 
profiles, participate in message boards, and form and join teams \cite{Anderson2004,Beberg2009}. Such an infrastructure 
stimulates individual and group competition. Empirical research suggests that team competition is particularly 
important. Being affiliated with a team significantly increases one's contribution. More importantly, being part of a 
team maintains high levels of contribution over time \cite{Nov2010}.

As PRC systems have grown in popularity and sophistication, new design challenges have emerged. These have been 
reflected in the structure of the machine-machine sub-network. One of the major issues in PRC is managing the expected 
high rates of machine node unavailability. In current PRC implementations, this has led to redundant computing 
\cite{Anderson2004,Mayer2015}. Redundancy also helps resolve erroneous computational results due to malevolent users 
or, more often, malfunctioning machines. Alternative solutions to node unavailability include designing a framework 
core architecture that contains an extra layer that monitors, manages, and brokers resources \cite{Cuomo2012} and 
designing a hybrid resource architecture by supplementing volunteer computers with a small set of dedicated, reliable 
computers \cite{Lin2010}.

Another major problem in current PRC systems is the back-end bottleneck due to the centralized storage of results. This 
problem can be resolved by combining PRC with public resource storage, thus limiting the volume of data transfer as well 
\cite{Beberg2009}. Additional issues of unique authentication, authorization, resource access, and resource discovery 
can be resolved by borrowing ready-made solutions from grid computing \cite{Foster2001}.

Finally, current implementations of PRC are limited to tasks with independent parallelism, meaning that participating 
machines are not required to communicate with each other \cite{Anderson2002}. Newer projects, however, take the form of 
peer-to-peer networks. For example, \citeN{Marozzo2012} develop a MapReduce-based peer-to-peer PRC system in which 
autonomous nodes are dynamically assigned slave roles, master roles, or backup master roles. In another example, 
\citeN{Mayer2015} implement a Cloud system of a loose set of voluntarily provided heterogeneous machine nodes that uses 
a gossip-style protocol for communication.

\subsection{Crowdsourcing}
\label{sec:crowdsourcing}

Jeff Howe originally coined the term crowdsourcing in the June 2006 issue of \textit{Wired} magazine to denote situations where 
a company or institution take a function once performed by employees, and outsource it to an undefined network of people 
in the form of an open call \cite{Howe2006}. The term ``crowdsourcing'' has since been used in a very broad sense to 
denote a range of HMNs \cite{Doan2011,Geiger2011}. Here, we use it to denote systems that are based on open calls 
for the voluntary undertaking of tasks \cite{Estelles-Arolas2012}, hence closely aligned with Howe's original 
definition. Crowdsourcing (CSO) is, as such, a top-down initiated process, and usually distinctly different from mass 
collaboration as in, for example, open software development (see Section~\ref{sec:masscollaboration}). 
The defining characteristics of these HMNs are that 
humans actively select tasks and contribute to them but do not generally interact and collaborate with each other directly. 
More importantly, humans rarely receive any feedback from the machine --- they do not obtain any information on or direct 
benefits from the project they contribute to (Figure~\ref{fig:types}B). In a sense, 
CSO platforms are similar to distributed computing systems: each user is equivalent to a 
processor that needs to solve a task requiring human intelligence \cite{Kittur2013}. 

Prominent examples of CSO HMNs include online microwork platforms such as Amazon Mechanical Turk and 
Crowdflower, voluntary mapping websites such as Ushahidi and 
OpenStreetMap, and citizen science projects such as Zooniverse
and FoldIt. Companies 
and organizations may also crowdsource innovation tasks to the general public (as with Threadless) or experts beyond 
the organization's boundaries (such as InnoCentive) \cite{Brabham2008}, 
a practice which is often utilized within the field of open innovation. 
Perhaps the CSO project that is most familiar to the general public is reCAPTCHA. Web users contribute to 
reCAPTCHA when they take a Turing test on a website in order to confirm that they are human and not a computer. As part 
of the test, they transcribe scanned words that optical character recognition programs failed to 
recognize and thus help digitize large volumes of old text.

The successful functioning of CSO HMNs requires recruiting and maintaining a pool of intelligent, diverse, 
and capable contributors \cite{Saxton2013} and managing and processing their contributions. 
Various approaches can be undertaken to successfully recruit and keep contributors. One way is to make the 
contributions implicit, as in reCAPTCHA. Most individuals complete the task unbeknownst to them --- they are simply 
trying to register for an online service or make a purchase, for example. Importantly, reCAPTCHA 
is neither more difficult nor more time-consuming than a regular CAPTCHA test
\cite{vonAhn2008}. 

Another method to motivate contributors is to remunerate them. Payments can be fixed or success-based. They can come in 
the form of a piece rate, a piece rate with a bonus, a quota rate, or a contest prize. Experimental research has shown 
that higher piece rates increase the quantity but not the quality of contributions; further, quota rates result in 
higher quality work for a smaller budget than piece rates \cite{Mason2010}. Prize-based contests are optimal for highly 
uncertain innovation problems as they involve high number of entrants who execute multiple independent trials, which 
increases the probability for a maximally performing solution \cite{Lakhani2013}. The effect of payment is not always 
straightforward, however, as it may reduce social motivation \cite{Frey2000}.

Much participation in CSO projects is voluntary and indeed driven by intrinsic and social motivation. To 
attract more volunteers, CSO platforms can emphasize the connection between people's contributions and the 
project outcomes. For example, citizen science projects can keep participants updated on the articles published as a 
result of their contributions \cite{Cooper2010}. Volunteer-mapping sites can successfully rally on social media around 
natural disasters \cite{Zook2010}.

Gamification is another strategy to recruit volunteers and encourage their prolonged engagement in a CSO 
project. A game offers the players a varied motivation set that leverages ambition, competition, and cooperation. Thus, 
a gamified CSO project should allow for short-term and long-term rewards through tracking game score and 
player rank, social praise by allowing comments on chats and forums, and collaboration by enabling the formation of 
teams \cite{Cooper2010}. These strategies have been successfully implemented in biology projects for discovering 
protein structures \cite{Cooper2010}, aligning multiple sequences of DNA \cite{Kawrykow2012}, and solving in vitro RNA 
design problems \cite{Lee2014}.

Managing the contributors' work poses the second major challenge in designing CSO platforms. Contributions 
need to be collected, processed, and aggregated. Although the quantity of contributions is what makes CSO 
work \cite{Surowiecki2005}, the quality of contributions remains of paramount importance. 

One major challenge with CSO is dealing with biases and errors in contributors' submitted work, whether malevolent or accidental. 
\citeN{Quinn2011} propose multiple techniques to do this: output agreement, input agreement, economic incentives, 
defensive task design, reputation systems, redundancy (combined with majority consensus), ground truth seeding, 
statistical filtering, multilevel review, expert review, and automatic check. \citeN{Allahbakhsh2013} add to the list 
run-time support, workflow management, worker selection, and contributor evaluation. The latter two have been 
implemented in sophisticated algorithms that assign weights to workers based on the quality of their contributions in 
order to account for biased responses \cite{Ipeirotis2010,Raykar2010} and eliminate responses by spammers 
\cite{Raykar2011}.

Beyond correcting for intentional or unintentional errors that have already occurred, one could implement more 
foresighted strategies to improve the quality of contributions. These strategies usually involve creating and 
structuring interactions between contributors. For example, \citeN{Kittur2011} and \citeN{Bernstein2010} implemented 
frameworks that manage the coordination between users to let them complete complex tasks such as writing and editing 
articles. \citeN{Silvertown2015} purposefully connected participants with experts in a hybrid learning-CSO 
platform for the identification of biological species. \citeN{Khatib2011} introduced tools for free collaboration 
between contributors that lead to the discovery of new algorithms for the folding of proteins. \citeN{Bayus2013} 
suggests that encouraging contributors to comment on diverse sets of ideas proposed by others could prevent creativity 
stagnation due to the ``fixation effect'' and help maintain an on-going supply of good new ideas in idea-generating 
CSO platforms.

\subsection{Web search engines}

A Web search engine (WSE) is a system that searches and returns content from the Web. Apart from formulating queries, 
users do not actively contribute to the system. Their contribution is passive: the WSE observes and analyzes their 
search behavior to improve and customize search results. Additionally, the search results are filtered and ranked, 
which is the most important function of these HMNs (Figure~\ref{fig:types}C). 

The greatest challenge that WSEs need to address is keeping up with the scale of content on the Web \cite{Arasu2001,Brin2012}. 
Since this 
problem entails complex technical solutions at the level of machine-machine interactions, it is out of the scope of the 
present survey. Instead, we here focus on design solutions related to the motivation of users to participate in the improvement of the output of WSEs. These design solutions are typically aimed at making user participation an 
implicit part of their search task as well as their linking behavior, which is reminiscent of the user motivation 
strategies of crowdsourcing HMNs such as reCAPTCHA discussed above.

Another key challenge of WSEs is providing users with relevant search results, quickly. Query assistance is one mechanism
introduced to help with this, which is addressing the direct interaction between the users and the WSE interface. 
One approach is to do this based on the user's own input of a query and subsequent query reformulations \cite{Huang2009}.
Alternatively, other users' inputs can be used by mining query associations from multiple users \cite{Carpineto2012}, which 
utilizes the HMN further.

To improve the search results, filtering and ranking can also use information about users' behavior. For example, 
the links between Web pages imply recommendations from users (owners of the respective Web pages) \cite{Arasu2001}. This
implicit notion is encapsulated in Google's PageRank algorithm, for example, as it determines the importance 
of a page based on the importance of the pages that link to it \cite{Page1999}. Another well-known ranking algorithm, HITS,
also uses recursive logic to differentiate two types of important Web pages, hubs and authorities: hubs are pages that point 
to many authorities and authorities are pages that are pointed to by many hubs \cite{Kleinberg1999}. Newer ranking algorithms 
tend to use even more information about the linking structure. For example, ClusterRank takes into account not only a page's 
number of linked neighbors and the neighbors' influence scores but also the neighbors' links with each other \cite{Chen2013}.

\citeN{Gollapudi2009} suggest that WSEs need to intentionally diversify results to improve user satisfaction; simply
providing users with the highest ranked pages is not considered sufficient. To personalize search results and improve their 
relevance, WSEs can also analyze individual behavior \cite{Steichen2012}. For example, \citeN{Collins-Thompson2011} propose 
algorithms to re-rank search results based on an estimation of the user's reading proficiency. \citeN{Jansen2008} suggest 
that WSEs can use knowledge of user intent (whether informational, navigational, or transactional) to provide more relevant 
results.

Finally, WSEs can also use peers' behavior to improve the results from individual searches. This relatively recent 
model is known as social search. First, WSEs can potentially incorporate search data from social bookmarking 
\cite{Heymann2008}. Second, WSEs can mine the profiles of a user's online social network friends and return results 
based on their recommendations \cite{Morris2010}. WSEs can also be adapted to support collaborative search in real 
time \cite{Morris2007,Morris2013}.

\subsection{Crowdsensing}

In crowdsensing (CSE) applications, users with computing and sensing devices (for example, smartphones or fitness trackers) 
share data and the application uses the data to measure and map certain phenomena \cite{Ganti2011}. The contributions 
by users can be active, as well as passive. The contributions are then analyzed and action and behavior recommendations 
are sent back to the user (Figure~\ref{fig:types}D).

CSE has been employed in many different areas, including traffic, health, and environmental monitoring 
\cite{Khan2013}. Applications have been developed to create real-time awareness of earthquakes \cite{Faulkner2014}, 
track and encourage physical activity and healthy lifestyle \cite{Consolvo2008}, track personal transportation patterns 
to encourage ``green'' transportation behavior \cite{Froehlich2009}, and predict bus arrival times \cite{Zhou2012}, 
among many others.

Participating in CSE incurs real costs: increased consumption of energy and bandwidth, as well as increased 
risk of leaked personal information \cite{Ra2012}. As a result, most research on CSE has focused on designing 
monetary incentives for participation, reducing participation load, improving energy efficiency, and guaranteeing 
privacy. 

To guarantee an adequate number of participants, some researchers have proposed monetary incentives through reverse 
auctions, whereby users claim bid prices for their sensing data \cite{Koutsopoulos2013,Lee2010}. Others have suggested 
that this could be achieved through opportunistic sensing, that is, sensing that is fully automated and does not 
require the user's active involvement \cite{Lane2010}. Full automation, however, is difficult to achieve technically, 
as the application needs to combine data from multiple sensors to infer the context. Instead, others have focused on 
improving the energy efficiency of the CSE applications. \citeN{Lu2010} present methods to turn sensing on and 
off depending on the quality of input data and the user's long-term behavior and mobility patterns. \citeN{Sheng2012} 
propose a collaborative sensing approach, whereby data collected from mobile phones is analyzed in real time on servers 
in a Cloud in order to calculate the best sensing schedule and inform the phones when and where to sense. 

Guaranteeing privacy is a more challenging problem to solve. CSE applications are in danger of leaking personal 
data such as users' location, speech, potentially sensitive images, or biometric data \cite{Christin2011}. Even if 
sensing data look safe, they may be reverse-engineered to reveal invasive information. Last but not least, there is 
also the ``second hand smoke'' problem: a person with a sensor can undermine the privacy of nearby third parties 
\cite{Lane2010}. 

Effective privacy-protection measures can be implemented at every stage of the path of the data, from collection to 
consumption \cite{Christin2011}. When reporting data, spatial cloaking techniques can be combined with data 
perturbation to improve user anonymity without affecting aggregate data estimates \cite{Huang2010}. During that stage, 
phones can also combine their data with data from their neighbors before transmitting it to the application server 
\cite{Li2014}. Techniques like this can be effectively combined to allow even for anonymity-preserving reputation 
systems \cite{Wang2013}. However, it has been shown that even if the collected temporally and geographically tagged 
data are anonymized, only few data points can be uniquely linked to individuals in large crowds \cite{Montjoye2013}.  

Processing and analyzing the collected data presents the second substantial design challenge for CSE systems. 
Due to user mobility, density, and privacy preferences, the data delivered often have many measurement gaps in both time 
and space \cite{Ganti2010}. To be able to generalize from a sparse sampling of high-dimensional spaces, 
CSE systems can employ sophisticated data interpolation techniques \cite{Mendez2013}. 
Further, to make best use of the data a 
user reports, they can infer the context with the help of machine learning techniques 
\cite{Pejovic2015}. Finally, they can combine context-rich data from multiple users to predict models of future behavior 
and offer action recommendations \cite{Pejovic2015}.

\subsection{Online markets}

Under the type ``online markets'' we group a number of HMNs that involve the exchange of goods and services. 
The things that are exchanged differ but what unites these HMNs is the common relational structure. In particular, 
users can view each other's contributions and, in some cases, evaluate them, but cannot modify them. More 
importantly, while users actively make contributions, the service does not filter or rank them but simply lists them, 
either individually or in an aggregate form (Figure~\ref{fig:types}E). In other words, 
the service usually does not bias the content a user is exposed to. 

Online markets are very familiar to most Web users. First, there are consumer-to-consumer markets such as Craigslist, 
eBay, Uber, and Airbnb, which connect users who offer material goods and services with users who seek those goods and 
services. File sharing networks such as Gnutella and BitTorrent are similar with the exception that the exchanged goods 
are digital. Finally, prediction markets, such as the Iowa Electronic Markets and TradeSports, where users trade 
``stocks'' tied to future events, are another prominent example of online markets. 

Peer-to-peer (P2P) and file sharing markets involve unregulated direct inter-user transactions and as a result, their 
functioning crucially depends on trust. The problem of trust online is most commonly addressed by designing and 
implementing reputation systems \cite{Josang2007}. In a reputation system, users rate each other after a transaction 
and the aggregated ratings inform other users in their own transaction decisions. When designing reputation systems, it 
is important to take certain well-established empirical facts about social influence in consideration. Previous 
research has shown that when users deal with unknown sellers or providers, positive reputation matters but negative 
reputation is much more decisive \cite{Standifird2001}. The opposite appears to be true when users interact with 
well-known brands: positive information has a stronger effect on purchasing decisions than negative information 
\cite{Adjei2010}. High reputation means high trust by others but who those others are also matters \cite{Baek2012}. For 
example, recognition by a well-known other, such as an institution, affects one's trust positively to a higher extent 
than recognition by one's peers \cite{Jones2008}. The idea to weigh reputation by the trustee's importance in the 
trusting network has been implemented in two recent reputation algorithms. The EigenTrust algorithm \cite{Kamvar2003} 
is based on the rule of transitive trust: if I trust a user I am also likely to trust the people that that user trusts. 
The PowerTrust algorithm \cite{Zhou2007} relies on the idea of identifying the few power users who can indisputably 
serve as authorities. 

Individuals' estimated trust and reliability could also inform structural re-designs of the online market. For example, 
\citeN{Saroiu2001} propose to use this information to delegate different responsibilities to different users in P2P file 
sharing applications. Such intervention is intended to mitigate free riding, where most users consume without paying 
back to the community in return \cite{Hughes2005}.

Information on existing social structure can be used to improve not only trust and cooperation in online markets but 
also content search. \citeN{Sripanidkulchai2003} proposes to preserve the anonymity of the system but group 
users by interests (not externally visible). Since users who already share interests are likely to have more interests in common, a content 
location algorithm that exploits the interest overlay network results in faster content queries, lower system load, and 
improved scalability. Such implicit grouping by similarity in interests and behavior has already been implemented in 
``collaborative filtering'' recommender systems, which are common in online markets \cite{Schafer1999,Schafer2001}. In 
contrast, \citeN{Pouwelse2008} propose de-anonymizing the P2P system by introducing social-network capabilities in order
to improve content discovery and recommendation.

Prediction markets differ from P2P and file sharing markets because the exchanges in them are centrally regulated. 
Consequently, they are less affected by the problem of trust. Instead, the major problem they face is how to improve 
predictions. Once again, the solution has to do with careful design of the underlying user-to-user interaction 
structure. Prediction markets work because of the diversity of users \cite{Surowiecki2005}. This makes them robust to 
manipulation by a small group of individuals \cite{Hanson2006,Wolfers2004}. However, social influence can undermine 
diversity: it both shifts the average estimate and increases the users' confidence in it \cite{Lorenz2011}. Hence, 
prediction markets should reduce possibilities for interactions between users, such as observing users' current 
prediction, past performance, or formal expertise. Nevertheless, to improve the prediction market performance, it is 
possible to combine the predictions by the crowd with predictions by a panel of experts \cite{Prokesch2015}. 

\subsection{Social media}

In one of the most cited definitions, social media (SM) is defined as ``a group of Internet-based applications that 
build on the ideological and technological foundations of Web 2.0, and that allow the creation and exchange of User 
Generated Content" \cite[p. 61]{Kaplan2010}. Kaplan and Heinlein's definition is very broad, with the consequence that 
they include virtual worlds such as Second Life and mass collaboration projects such as Wikipedia as examples of social 
media. We use a more subtle definition and understanding, focusing on the typical interactions
among the nodes in the network. In SM, human nodes assess each other's contributions (rather than 
modify them) (see Figure~\ref{fig:types}F). SM is intended to enable users to form online 
communities and interact and share information in them \cite{Kim2010}. In SM, users can contribute content actively. 
They are also actively involved in observing and evaluating each other's contributions. The applications use these 
evaluations to filter the content that users receive. This filtering can be site-wide or user-specific. 

SM includes social network applications such as Facebook and LinkedIn. In these applications, users maintain a profile 
and a list of users with whom they are connected, both of which can be viewed by others \cite{Boyd2007,Donath2004}. SM 
also includes content communities such as Reddit and Tumblr, where users share news and media and vote and comment on 
each other's contributions. Some SM applications have the properties of both social network and news-sharing 
applications, e.g. Twitter \cite{Kwak2010}. Additional examples include review and rating applications such as Yelp and 
TripAdvisor and discussion forums and question-and-answer sites such as Yahoo Answers and College Confidential.

Participation in SM highly depends on users' motivation and concern for privacy. Based on data from Facebook users, 
\citeN{Lin2011} found that enjoyment is the strongest motivator to use SM. Still, utilitarian considerations play a 
role: users often use SM to organize social events 
and disperse news in an effective way \cite{Xu2012}. Further, SM is self-affirming, in the sense of satisfying the 
users' need for self-worth by allowing them to exhibit a successful, attractive, and well-connected version of 
themselves \cite{Toma2013}. More interestingly, there is a network externality effect on motivation: high levels of 
adoption of SM among one's peers not only increases the perceived enjoyment of the SM \cite{Lin2011} but also directly 
increases one's likelihood of adoption \cite{Sledgianowski2009}.

Protecting personal data and privacy is a complex problem in SM. Even if users keep their profiles private, their 
friendships and group affiliations sometimes remain visible. In addition, some of their friends may have public 
profiles. Previous research has shown that friendships, group memberships, and rating behavior can be used to infer 
sensitive personal attributes and information \cite{Kosinski2013,Zheleva2009}. Users' privacy is threatened by other 
untrustworthy users, ill-intentioned third parties, as well as the SM providers themselves. To protect users from other 
users, SM developers can analyze the SM network structure to infer users' trust \cite{Mislove2007}. Further, they can 
develop machine learning models to describe a user's privacy preferences towards each of his network contacts; these 
models can then be used to configure that user's privacy settings automatically \cite{Fang2010,Gilbert2009}. For even 
more strict privacy, the architecture of the SM application can be decentralized so that users store their private 
data on other users' machines, such as individuals whom they trust in real life \cite{Cutillo2009}.  

One of the most significant functions of SM is user-specific filtering and customization of content due to the sheer 
volume of data. While this addresses a practical issue from the users' perspective, it inevitably introduces
bias (intended or not) that influences the users. SM employs complex 
algorithms to predict the news that users would like to read, the media they would like to see, the people they would 
like to befriend, and the ads they are likely to succumb to. For example, SM applications can analyze user 
relationships to select high-quality content \cite{Agichtein2008}. They can also incorporate various behavioral 
information, such as locational co-occurrences, to recommend new links \cite{Scellato2011}. SM applications can take 
even more proactive roles. They can implement algorithms using the political valence of contributions to influence a 
user's opinion \cite{Bakshy2015}. They can filter content based on their emotional valence to sway a user's affect 
\cite{Kramer2014}. They can also encourage links that cut across traditional age and gender homophily in order to 
expose users to more diverse sources of information and influence \cite{Centola2015}. These strategies can help alleviate one of the major downsides of SM --- the potential for knowledge bubbles and misinformation epidemics due to selective interaction.

\subsection{Multiplayer online games and virtual worlds}

Multiplayer online games (MOGs) and virtual worlds (VWs) are simulated environments in which users 
participate and interact with each other and the environment via avatars that represent their identity. Users' actions
and contributions impact other users' experiences both directly and indirectly, by affecting the game world as a whole. 
In fact, in VWs, it is users who create the game world \cite{Ondrejka2004}. Apart from providing the interaction rules 
and settings, the MOG/VW service does not usually modify user contributions (Figure~\ref{fig:types}G).

The question as to why users repeatedly participate in MOGs and VWs has attracted much research attention. As with SM, survey 
studies have shown that providing optimal personal and social interactions improves users' experience and increases 
their loyalty \cite{Choi2004}. For MOGs, gratification from achievement and social interaction together with game 
incentives and fairness have been shown to make users more likely to continue playing \cite{Wu2010}. For VWs, perceived 
usefulness has been found to additionally increase the likelihood to participate \cite{Verhagen2012}. Moreover, the 
functional, experiential, and social motivations appear to be roughly equally important \cite{Zhou2011}. What is 
particularly unique for MOGs and VWs, however, is the sense of immersion or ``flow'' \cite{Yee2006}. Flow experience is 
the mental state of being fully absorbed and losing track of time \cite{Goel2013}. Providing optimal personal 
experiences and meaningful social interactions cause individuals to experience flow, which in turn increases their 
likelihood to continue interacting and playing \cite{Choi2004,Goel2013}. Customization (e.g., of one's avatar) further 
increases loyalty \cite{Teng2010}.

How does one design for meaningful social interactions in MOGs and VWs? Users may be involved in both negative 
interactions, such as attacks, and positive interactions, such as communication and exchange \cite{Szell2010}. Overall, 
however, observational research has shown that MOG users do not interact directly with other users as much as expected 
\cite{Ducheneaut2007}. This may be considered suboptimal because, in theory, MOGs and VWs can serve as a venue for 
informal sociability that fosters bridging social relations, which are relations that improve access to diverse 
information \cite{Steinkuehler2006}. \citeN{Ducheneaut2004} propose to force social interaction among users through the 
game design. For example, MOG developers can design interdependencies among characters and then design locations where 
these interdependencies can play out. Another solution to socially engage users is to design for indirect interactions. 
\citeN{Ducheneaut2006} argue that MOG users do not interact directly and collaborate with other users as much as use 
them as an audience, a source of entertainment, and a source of information and chatter. 

Designing a MOG/VW to keep users playing is a different problem from designing it to keep attracting new users. The 
more users there are, the more interactive and complex the virtual environment is. In order to guarantee continuous 
growth of the MOG/VW, one can reduce the learning and personalization costs in the short term \cite{Zhang2014}. In the 
long term, however, the problem of scalability looms large. A successful MOG/VW should provide consistent and secure 
service with fast response times for thousands of users simultaneously \cite{Claypool2006,Yahyavi2013}. Traditional 
client-server architectures, however, have inherent scalability limitations. In contrast, P2P architectures can achieve 
high scalability at low infrastructure cost \cite{Hu2006,Yahyavi2013}. In the context of MOGs and VWs, a P2P architecture 
means that each user's machine may hold master copies of some of the game objects and be responsible for propagating 
updates to other nodes. While improving scalability, such an architecture is vulnerable to cheating, security attacks 
and churn, problems that remain outstanding design challenges today \cite{Yahyavi2013}. 

\subsection{Mass collaboration}
\label{sec:masscollaboration}

Compared to the other HMNs, mass collaboration networks involve highly intense interactions 
among humans and machines in the context of a ``project'' set up for a specific purpose (Figure~\ref{fig:types}H). 
They require the highest level of involvement from users in terms of time and effort. Users can modify and reject each 
other's contributions and affect the project as a whole. Their contributions and the project content are often 
monitored and adjusted by the project leaders. This centralized oversight can happen automatically, for example, 
through vandalism-detection algorithms. Nevertheless, often, the organization of these HMNs is bottom-up. 

Wikis and open-source software (OSS) projects present two of the most prominent examples of mass collaboration HMNs. 
Wikis enable the simultaneous collaborative creation, modification, and deletion of content \cite{Tapscott2011}. Wikipedia, 
the online encyclopedia is the 
largest, most successful, and most popular wiki project. OSS projects allow the collaborative development and free 
distribution of computer software. Examples include the Firefox web browser, the Apache HTTP Server, and the Linux operating system.

Mass collaboration projects are developed by geographically and organizationally dispersed contributors, most of whom 
are volunteers. As a result, recruiting and retaining contributors is a critical design issue for OSS and wiki 
communities. \citeN{VonKrogh2003} suggest that ``joining scripts'' are important if users want to gain access to the 
collaboration community. These ``scripts'' are implicit constructs that determine the typical level and type of 
activity a joiner needs to go through before becoming a contributor. Users often decide to join and participate in a 
collaboration project because they have pragmatic considerations or expect external rewards. Survey results have shown 
that external rewards have greater weight for participation than internal factors, such as intrinsic motivation, 
altruism, and community identification \cite{Hars2001}. In fact, a high number of developers are paid for their OSS 
development efforts; others use their participation to improve their own software; still others receive benefits in 
terms of reputation and self-development. Contributors are able to learn even from mundane tasks such as reading and 
answering users' questions \cite{Lakhani2003}. In addition, contributors can benefit by exchanging valuable work with 
each other.  A theoretical model has shown that since more modular codebases with more option value foster such 
exchanges, they increase recruitment and retention and decrease free riding \cite{Baldwin2006}.

Still, external rewards are not the sole motivator. First, they appear to be less important to wiki contributors, 
compared to OSS contributors \cite{Oreg2008}. Second, they appear to be only driving the decision to join but not the 
decision to stay \cite{Shah2006}. Sustained participation appears to be better predicted by group identity and 
community belonging. For example, \citeN{Hertel2003} show that self-identification as a Linux developer is one of the 
factors that determine engagement in the Linux project. \citeN{Fang2009} similarly find that situated learning and 
identity construction are positively linked to sustained participation, at least in the phpMyAdmin OSS community. 
\citeN{Bagozzi2006} confirm that active participation is associated with group-referent intentional actions.

In addition to recruiting and retaining contributors, mass collaboration platforms need to organize the leadership, 
coordination, and collaboration among the contributors \cite{Crowston2012}. Analyses of the e-mail communication 
network of contributors to several OSS projects reveal that sub-communities emerge naturally to mirror collaboration 
relations \cite{Bird2008}. The community further subdivides into a core of usually 10-15 developers who create about 
80\% of the code functionality, a much larger group around the core who repair defects in the code, and an even larger 
periphery of users who report problems \cite{Mockus2002}. Members of the core usually know and trust each other and 
communicate intensely to manage the dependencies among the contributed code.

How do these naturally emerging structures 
affect the growth and the success of the OSS/wiki project? \citeN{Crowston2005} analyzed 230 project teams on SourceForge 
to find that larger projects tend to have more decentralized communication networks. Based on a longitudinal analysis 
of an-order-of-magnitude larger SourceForge sample, \citeN{Singh2008} conclude that internal cohesion, as defined by 
repeat ties, third-party ties, and structural equivalence among contributors, is also associated with project success. 
Similarly, \citeN{Hahn2008} find that a project is likely to attract more developers if prior collaborative relations in 
the OSS developer network exist. However, contributors' external embeddedness has more complex effects on success. 
While a high number of external contacts increases success, only moderate technological diversity of the external 
network and moderate external cohesion are beneficial \cite{Singh2008}. Further, contributors' participation in 
multiple projects can both improve and aggravate the project's chances for technical success \cite{Grewal2006}.

\section{Discussion}
\label{sec:discussion}

Based on the HMNs reviewed above, here we provide further discussion 
on the emerging trends, common challenges, design implications, and future considerations.

\subsection{Emerging trends}

Our survey suggests five major trends that concern most of the HMNs.

First, human-human interactions appear to be intensifying in HMNs. Many systems are starting to allow for social 
interactions among their users. In addition, many of them are looking for ways to make the social interactions more 
immediate and intense. In a sense, everything can now become a social network; everything can be ``shared.'' 
Social events and meet-ups have been shown to be effective in recruiting new users as well as stabilizing the commitment 
of pre-existing users \cite{Hristova2013}. Social trading, based on text-based chats 
among the traders in a company, has also been introduced recently \cite{Saavedra2011}. Both of these cases are examples of 
machine mediated collective behaviors in which social interactions are not included in the original design. 

Efforts towards strengthening social interactions have been evident in numerous other research areas 
as well. For instance, \citeN{Skowron2014} discuss two studies conducted with an affective dialogue system in which 
text‐based system-user communication was used to model, generate, and present different affective and social interaction 
scenarios in order to intensify human-human interactions within HMNs. In the same framework, the SoCS project aimed to
support higher quality online deliberation in HMNs, especially by supporting a number of ``social deliberative skills" 
such as perspective taking, empathy, self-reflection, tolerance for uncertainty, listening and question-asking skills, 
and meta-dialogue in online contexts \cite{Murray2014}.

Second, and related to the first trend, human-machine interactions are becoming less demanding. By introducing social 
functionality, HMN designers capitalize on social motivations to encourage participation. In some cases, HMN designers 
have also started introducing monetary incentives, as in paying for contributions on review and rating sites or for 
crowdsensing applications. Last but not least, automated data collection is gaining prominence. These changes serve to 
reduce the effort on behalf of participants and increase their gains from participation. HMNs no longer need to rely on 
``altruists'' to continue functioning and growing. Previous research has shown  
that the introduction of social functionality in HMNs eases their overall operation and node interaction. 
For instance, \citeN{Murray-Rust2014} present models and techniques for coordination of human workers 
in crowdsourced software development environments. The techniques augment the existing Social Compute Unit concept 
-- a general framework for management of ad-hoc human worker teams. This approach allows the researchers to combine 
coordination and quality constraints with dynamic assessments of software-users' desires, while dynamically choosing 
appropriate software development coordination models, leading to easier human-machine interactions.

Third, machine-human interactions are also becoming more prominent. Machines now customize and filter any information 
that users receive and consume. This is evident in numerous HMNs nowadays. A typical example is Facebook and the content 
control carried out within this network. Even in such a massive network, advanced content 
control algorithms are applied and have proven to be quite effective \cite{Bakshy2015}. 
Similar algorithms are applied in several other 
HMNs enhancing machine-human interactions.  

Fourth, not just the nature of interactions but also the actual structure of the HMNs is 
starting to evolve. There is a trend towards redesigning HMNs as peer-to-peer networks. A typical example regarding that 
trend is Bitcoin. The Bitcoin cryptocurrency is built on top of a decentralized peer-to-peer (P2P) network used to 
propagate system information such as transactions or blockchain updates. The Bitcoin architecture does not rely on a 
centralized server. Instead, a distributed approach has been adopted to support the system \cite{Donet2014}. Such an 
approach can be used in many of a system's facets, including data storage, data confirmation, and data transmission.
The major reason for this is that P2P architectures improve the scalability of HMNs. But they also entail certain 
undesirable consequences. In P2P networks, machines become equivalent with users, as each user contributes a machine to 
store, process, and transmit data. This means that machines start to exhibit some of the problems that humans have: 
unavailability, unreliability, distrustfulness, and untrustworthiness. This implies that social science knowledge and 
approaches are starting to be indispensable even for the engineering of HMNs.

The fifth trend is a product of the previous four --- the number of HMNs that cut across the eight types 
we have identified is starting to increase. For example, cryptocurrencies such as Bitcoin have the properties 
of both public-resource computing and online markets. Online and mobile dating services such as Match.com and 
Tinder have characteristics from both online markets and social media. Wildlife@Home, a project aimed at analyzing 
videos of wildlife, combines public resource computing with crowdsourcing \cite{Desell2013}. 
Open innovation projects often extend the crowdsourcing approach to mass collaboration, by combining 
ideation with collaborative refinement of ideas. Opportunistic Internet of Things systems combine crowdsensing 
with social media \cite{Guo2015}. Since these hybrid types of HMNs are relatively new, 
the literature on them is still growing. Our survey methodology placed an emphasis on publication impact and as a result, 
down-weighted these networks. Nevertheless, this trend of hybridization renders our survey particularly valuable, 
as it emphasizes the need for the transfer of knowledge and design solutions across different types of HMNs.

\subsection{Common challenges}

In this survey, we analyzed different types of interactions to identify 
different types of HMNs. However, there are risks and challenges in relation to the interactions that are common among all HMNs. 
Here we discuss issues related to user motivation, trust, user experience, privacy, and scalability.

Attracting users and keeping them motivated to participate and contribute is a major challenge across all identified HMNs.
A number of approaches to mitigating this challenge
have been proposed, as we have seen in Section~\ref{sec:networks}. A basic distinction 
between these approaches concerns whether to exploit extrinsic or intrinsic motivational factors, 
and different types of HMNs exploit such motivational factors differently. HMNs for public resource computing and 
multiplayer online games typically aim for intrinsic participant motivation, for example by targeting the civic 
benefit or experiential value of the HMN. HMNs for crowdsensing and online markets, on the other hand, typically 
address recruitment and motivation challenges through extrinsic motivational factors such as remuneration or utilitarian 
gains. Possibly, when designing for increased user recruitment and motivation for one HMN type, it could be beneficial 
to look to other HMN types for inspiration. For instance, design for participant recruitment and motivation in 
crowdsensing systems could perhaps to a greater degree accentuate the civic value of participation and, hence, to a 
greater degree benefit from intrinsically motivated participation. To make such intrinsically motivated participation 
practically feasible, however, participation costs would need to be reduced. One way is to follow
the example of crowdsourcing 
initiatives such as reCAPTCHA where the contributions are made as integrated parts of users' everyday behavior 
\cite{vonAhn2008}. Another example of making data collection an integrated part of users' everyday behavior is the use of 
mobile phone network position data to monitor traffic flow \cite{Calabrese2011}. Such integrated approaches to 
data collection could possibly also be applied in crowdsensing.

Other approaches to participation recruitment and motivation are more specific to particular HMN types. For example, 
the use of informal joining scripts \cite{VonKrogh2003} has been shown to be an efficient approach of relevance to 
establish participant motivation in systems for mass collaboration, in particular as existing collaborators need to 
filter out non-serious newcomers. However, for other HMN types, such as online markets or crowdsourcing, a more 
adequate approach for the same purpose may be to apply traditional recommender systems \cite{Schafer2001}, as 
participants here typically connect in multiple brief engagements and informal joining scripts are therefore neither 
effective nor efficient.

Trust is another major challenge for HMNs. Some participants 
may struggle with a concern that the technology will undermine their own position or may be unwilling to rely 
entirely on the capabilities of the machines \cite{Lee1992,Lee2004}. Others may be mainly concerned with the security of 
their private data \cite{Dwyer2007}. In response, machines need to be transparent and predictable, to create a 
perception of trustworthiness \cite{Cheshire2011,Corritore2003,McEvily2003,Schoorman2007}. On the other hand, they may 
also need to identify and counteract deliberate malevolence: hacking, vandalism, and intentional abuse of their 
algorithms, such as Google bombs and link farms.

Machine interventions to the content that observers see and use can lead to undesirable self-reinforcing feedback loops 
that can negatively impact the service rendered. For example, if the number of users who see a particular contribution 
increases the chance of others seeing it, the diversity of content can decrease. From the participants' point of view, 
there is again the problem of trust, since the machines in HMNs often employ ``black-box'' algorithms that can affect 
one's behavior and mood without the participant's knowledge and explicit consent \cite{Bond2012,Kramer2014}. These two 
related effects have been labelled ``filter bubble'' \cite{Pariser2011}. However, such filtering or adaptation of 
content may be both necessary and beneficial. Without such filtering, the result may be information overload on the 
humans in the HMN. Further, from a machine network perspective, the size and amount of content to be delivered will 
impact network performance and therefore potentially undermine the Quality of Experience (QoE) for the user. This may 
lead to the monetisation of content delivery, imposing strict controls and monitoring of throughput especially in 
last-mile segments.

A concern for QoE in machine-human interactions is closely related to Quality of Service (QoS) between machines within 
a network supporting HMN capabilities. In traditional plain old telephone service (POTS) networks, this often came down 
to routing and providing redundant configurations to ensure that point-to-point communication could be guaranteed 
irrespective of potential failures or bottlenecks. HMNs similarly require complex assemblages of multiple computers and 
devices; in some cases dedicated links and subnetworks are required to guarantee QoS at least (such as Content Delivery 
Networks). The machines communicate via Bluetooth, the Internet, or satellite, depending on proximity and the required 
QoS. Since not all such machines will be owned by the same service provider or carrier, contractual arrangements must 
be in place in support of agreed levels of service. As and when these are in place, there is also potentially the issue 
of transmission speeds and throughputs, leading to dedicated networks, packet-inspection and/or local caching. 
Although justified on the basis of service delivery, this could compromize fairness as well as privacy 
\cite{Gedik2008,Julian2002,Kamble2004}. With privacy and data sensitivity in mind, there may be a need to impose a 
logical network on top of the existing structures (Virtual Private Network), requiring specific security 
protocols agreed between the different components of a network. Security architectures often mimic interpersonal 
trust relationships in this way, effectively introducing trust transfer into a purely machine-based exchange 
\cite{Hardjono2004,McEvily2003,Stewart2003}. In other cases, often in connection with issues of robustness as well as 
privacy, dedicated networks, including private and reserved broadcast wavelengths will be used to carry data for 
specific purposes or in special cases \cite{Chiti2008,Salkintzis2006}. There may even be specific automated practices 
introduced to impose security compliance without human intervention. 

Another significant challenge faced by most HMNs is scalability \cite{He2011,Ugander2013}. As the number 
of humans in the HMN grows, 
machines need to connect, communicate, and compute in a more efficient way, including sometimes sophisticated 
routing and resource allocation management to ensure connection between the right user(s) and the right machine(s) 
\cite{Baran1989,Broch1998,Cardellini1999}. In addition, they need to be able to provide consistently good bandwidth, 
guarantee compatibility between different communication protocols and operating systems, and improve data structures 
for efficient querying, caching and storage \cite{Chen2008,Mehlhorn1989}.

\subsection{Design implications}

Our framework helps identify common design approaches to the challenges that different HMNs share. We can group the design implications according to the four types of interactions we investigated: human-human, human input to machine, machine output to humans, and machine-machine.

Regarding human-human interactions, the major design concern is how to recruit human actors and encourage productive or constructive interaction between them. To secure a solid base of users, special attention should be given to attracting new users through peer influence and protecting them from bouncing by streamlining their first contributions and preventing harassment by more experienced users. Further, design solutions should be geared towards improving engagement and collaboration between users and creating a feeling of community. This could be achieved by, for example, organizing offline social events, symbolically rewarding users with badges, applying a loyalty ladder, strengthening within-platform communication, and strengthening trust through rich profiles and recommendations. 

The design implications concerning human input are more typically associated with the User Interface (UI) and the more traditional HCI domain. They relate to encouraging as well as facilitating human interaction. Human actors need to be encouraged to use the interface and to behave in ways that provide the types of input required. This may be associated with the persuasiveness of the interface, though this is not the whole story. Continued use requires efficiency and perceived usefulness, as captured in the technology acceptance model \cite{Venkatesh2000}. Further, trust and privacy should be ensured, for example, by offering strict and clear privacy policies and employing transparent algorithms. Additional design solutions include organizing contributions around campaigns rather than routine as is often done with crowdsourcing projects related to natural disasters and encouraging shared responsibility for the HMN by requesting regular feedback from users on the the overall state of the project. 

The design issues regarding machine output concern how machines respond to humans and what they do with the information and content provided. It is important that machines respond proactively and are perceived as positive contributors. Design solutions should consider not only advanced content filtering functionality to address information overload but also algorithms to detect original and emerging content beyond popular authoritative sources. Algorithms can route content and tasks to users according to their interests and expertise in order to improve the quality of contributions and retain contributors. Further, algorithms can sift through content to select high quality contributions to show as examples to users in order to improve their future contributions. Machines can also creatively employ the geographical distribution of the HMN users to encourage the formation of local collaboration groups, distribute work load or even server load \cite{Ugander2013}, and tackle projects and tasks that are inconceivable otherwise. For example, projects that require continuous input can be intelligently routed to users from different time zones.

Regarding machine-machine interactions, it is not enough for human actors to perceive the effectiveness of the machine components they are most clearly involved with, that is, their own computers. Instead, for the HMN to be a success, there must also be a correspondingly productive interplay between the machines themselves as effective participants. Security is of primary concern here. When machine agents in a network wish to interact with others, often unknown to them, there needs to be a separate security negotiation including authentication and authorisation. One way around this would be introduce a single ``authority'' who would vouch for individual machine agents to mediate their access to other services.

\subsection{Future considerations}

Most previous work on networks has tended to focus either on social networks as communities of people 
sharing and interacting with one another online or offline, or on machine networks in terms of their robustness and complexity. 
It is our view, however, that in combining the two, we begin to appreciate how human-machine interactions can and 
often do lead to subtly interdependent behaviors which would have been impossible before.  Most earlier analysis 
of networks and behavior has tended to treat the actors or entities within the network as 
equivalent within their defined type. Moreover, the characteristics and behavior of those entities were seen to remain fixed and 
predictable once defined \cite{Boccaletti2006}. In order to develop a more advanced and comprehensive analysis of 
HMNs, we need to consider that the exhibited characteristics of nodes are dependent 
upon context, which may change in a complex way in response to a number of different factors \cite{Milo2002,Newman2003}. 
For example, human actors may be described as in-group and out-group, which defines their behavior in relation 
to other connected humans \cite{Giles2012,Hogg2006}. Similarly, a human actor may take the role of leader (or ``seed'') 
at one point in time, but may relinquish that role in favor of other actors in authority at a later point in time 
\cite{Aguirre1998,LemonikArthur2013}.

Machines are, as ever, more predictable than humans, but it is still conceivable 
that they may take more than one role in a network --- for example, a mobile phone may be a communication device,  
a display device, or a sensor under different circumstances or at different stages of an evolving network 
situation \cite{Arminen2006,Raento2009,Vinciarelli2012}. Each entity in the network can be characterized by a number 
of attributes, some of which are intrinsic to the entity (such as patient, doctor, collaborator, contributor) and some 
of which are context sensitive (such as leader, seed, follower, in-group, out-group, passivity) \cite{Horwitz2004,Newman2003}. 
Similarly, the relationship between entities can be characterized as a function of both context and time --- so trust 
and reliance are an attribute of the evolving relationship between two entities 
\cite{Rousseau1998,Schoorman2007,Sollner2012,Sollner2013}, whereas trustworthiness is an attribute of a single entity 
fulfilling a specific role \cite{Corritore2003,Rousseau1998}. These attributes of entities should be identified 
and characterized, their potential to adopt different behaviors and likelihood to adapt or respond to unexpected 
interactions, as well as the time-variant or dynamic nature of the relationships should be incorporated 
\cite{Milo2002,Newman2003}. 

The connections between entities in a network, edges, are often taken simply to represent the existence 
of a relationship, effectively taking the form of an infinite capacity conduit \cite{Boccaletti2006}. However, 
the edge itself may also display characteristics that influence behavior in the network. The edge is effectively a 
path across which information flows between two entities, but it can influence both the type of interaction and the 
perception by a human entity of another machine or human node \cite{Lusseau2003,Newman2003}. For example, 
the connection may be a slow or unreliable link, it will have a capacity limit which determines the type 
of information it can carry (text, data, audio, images) and as such may influence the level of trust or reliance 
between nodes and may therefore affect the response that a person has to the information available to 
them \cite{Lu2007,Newman2003a}
. Future analysis therefore needs to take into account the intrinsic characteristics of 
the edge as well as relate it to other network characteristics that may be influenced by it.

In reality, the emergence of new behaviors in an HMN is very difficult to predict, although 
the effect that new user behavior has on the operation of a network can be profound \cite{Jones2015,Lusseau2003}. 
As an example, consider the impact of social media usage on networks of people mediated by smart phones. 
The shift in behavior from verbal communication (telephony) to text communication has driven the developing 
capabilities of smart phones and the demand for universal wide bandwidth data connection. The inclusion of cameras 
of ever-increasing resolution and sophistication and biosensors in smart phones, and the ubiquity of apps has 
resulted in behavioral change in the users of such devices and networks. In social media networks such as Twitter 
and Instagram the emergence of the use of hashtags has revolutionized the sharing and discovery of information 
and has led to the adoption by the network of concepts such as ``trending'' 
\cite{Brock2012,Doctor2013,Fortunato2013,Glasgow2013}. These examples are indications of the way in which 
human users of any technology will adapt it to their own requirements, often beyond the expectations of 
the originator of the technology, and once adopted and widespread the technology provider will then move 
to incorporate the new usage into their offering. This is another aspect of the dynamic nature of a 
technology-based network.

That networks evolve has been established above, along with the reasons for their evolution. Any characterisation of a 
network therefore needs to embody the dynamic nature of the entities and connections between them, particularly in 
relation to the changing context of both the individuals and the network as a whole \cite{Fortunato2010,Milo2002}. 
Moreover, the underlying capabilities of the network will evolve with time, possibly leading to ``tipping points'' 
at which new or emergent behavior becomes possible, and this is paralleled by the developing competence of the 
network users whereby they are able to use the technology more effectively \cite{Boyatzis2006,Rogers2008}. 
The very existence of the network and the sharing of experience between people using the network ensure that 
behavior can become ``viral'' and spread rapidly between users \cite{Bakshy2012,Kietzmann2012}. This is not just 
``herd behavior'', although it can lead to such phenomena \cite{Langley2014,Ronson2015}, but can also be part of 
a shared learning phenomenon, which can ultimately lead to modifications in the way that the technological 
components of the network are implemented \cite{Benbya2006}.

\section{Conclusion}
\label{sec:conclusions}

This survey presented the state of the art in the field of Human-Machine Networks (HMNs). We focused on the actors in these 
networks and the interactions between them. We identified eight types of HMNs: public resource computing, 
crowdsourcing, web search engines, crowdsensing, online markets, social media, multiplayer online games and virtual 
worlds, and mass collaboration. These types differ in the structure and intensity of the interactions among and between 
humans and machines. We systematically collected relevant research on these types, with an emphasis on recent and 
high-impact work. We reviewed this work with a focus on issues related to designing the HMNs: motivating participants, 
guaranteeing their privacy, designing reputation, recommendation, and content-ranking algorithms, aggregating and 
processing contributions, and so on.

We have identified and discussed five emerging trends in HMNs: human-human interactions are intensifying in HMNs; 
human-machine interactions are becoming less demanding; machine-human interactions are becoming more prominent; the
structure of HMNs is evolving; more hybrid types of HMNs are emerging. 
The five development trends suggest that the differences 
between the eight HMN types we identified are starting to 
blur. Nevertheless, our analytical framework remains useful for identifying specific niches for development and 
innovation. Some possibilities include citizen science projects that encourage collaboration between users, search 
engines that rely on users explicitly ranking sites, and social network sites that enable file sharing.

Our framework helped us identify shared challenges, such as those related to 
user motivation, trust, user experience, privacy, and scalability. 
Our framework also helped organize the design implications from these challenges. 
An avenue for further work is to more systematically delineate types of HMNs 
and characteristics relevant to describing them in order to identify transferable solutions that may benefit
ICT developers when designing new HMNs. Building upon the survey presented here, 
\citeN{Eide2016} have already taken first steps towards a more systematic typology.

Key to our definition of an HMN is the synergistic effect that can occur from the interactions between 
humans and machines. Synergy may result from emergent behavior, which can be challenging to predict. We have discussed 
several aspects of emergent behavior and network dynamics pertaining to, e.g., context of use, changing role in the network,
and type of connections between actors. Characterisation of an HMN needs to embody the dynamic nature
of the actors and the connections between them. How to characterize HMNs to allow for the prediction of emergent behavior
is another interesting avenue for further work.

Our framework has its intrinsic limitations as well. We have simplified the type of interactions 
and limited the possible combinations. Moreover, we only considered few modes of contributions by actors, 
and we did not allow the actors to take different roles. Still, the simplicity of 
the provided framework allows for straightforward extensions in future iterations. 



\bibliographystyle{ACM-Reference-Format-Journals}
\bibliography{hmn}

\received{November 2015}{March 2016}{June 2016}

\elecappendix

\medskip

\section{Procedure for the Systematic Literature Search}

In this appendix, we describe how we executed the systematic literature search. The first step
was to identify specific types of HMNs. We did this by 
brainstorming for concrete examples and describing the interactions among humans and machines. This resulted in 
differentiating between three kinds of interactions (in addition to ``no interaction'') for 
each of the three directed dyadic relations we investigated (human-human, human input in human-machine interactions, 
and machine output in human-machine interactions). The grouping of the HMNs resulted in eight types, out of the $4^3 = 
64$ possible.

\begin{figure}
\centerline{\includegraphics[scale=0.65]{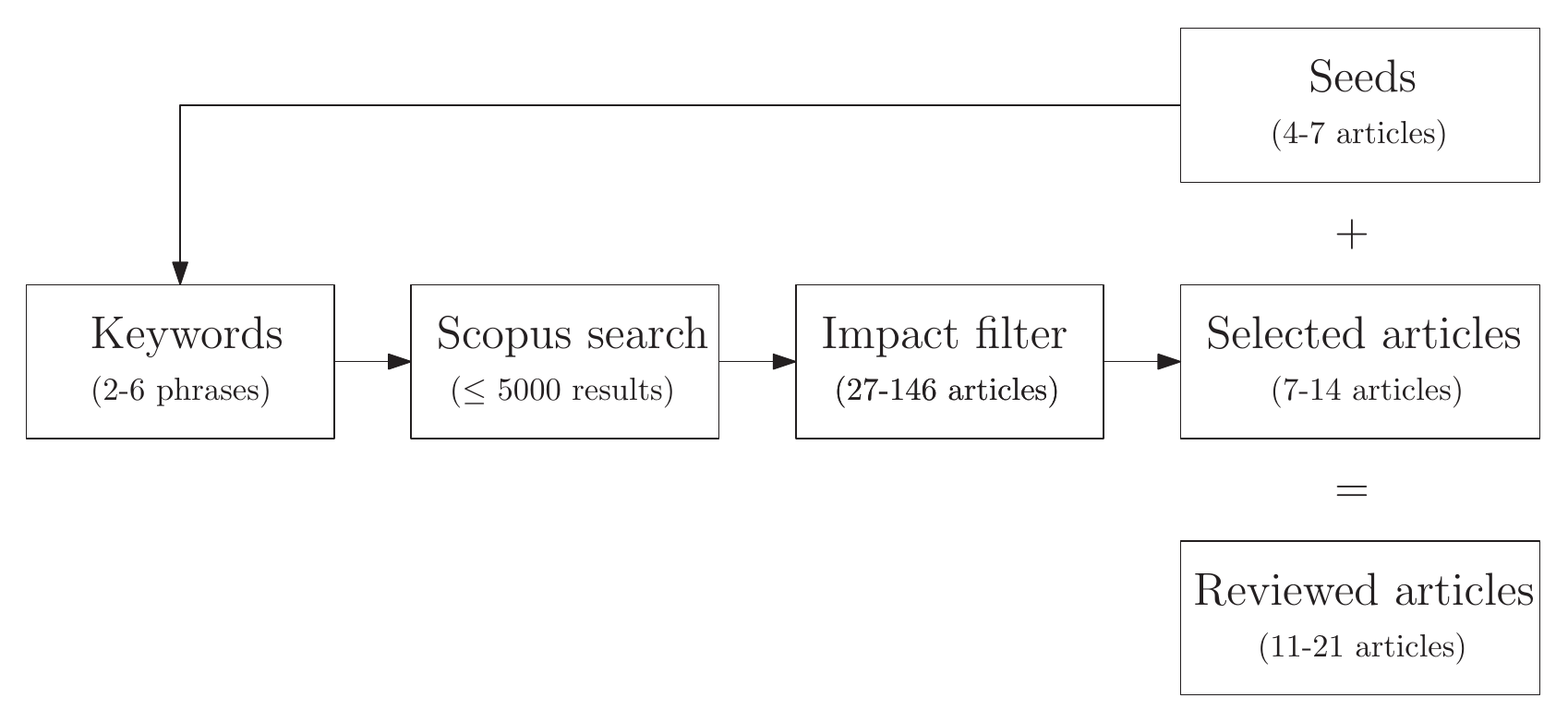}}
\caption{Method for the systematic literature search.}
\label{fig:method}
\end{figure}

Second, we used the eight HMN types to systematically collect relevant literature (illustrated in 
Figure~\ref{fig:method}). We started by identifying 4-8 ``seeds'' in each of the types; to represent
the most relevant articles with the highest impact. We conducted exploratory keyword search on Google Scholar for each 
type and selected the articles with the highest number of citations, with an intentional bias for literature review 
articles. Due to the emphasis on number of citations, there was also a de-facto bias for older publications. We then 
reviewed the content of the seed articles to manually compile lists of search keywords for each HMN type 
(Table~\ref{tab:keywords}).

The lists of keywords were used for systematic searches within the titles, abstracts, and keywords of all articles on 
the Scopus bibliographical database (www.scopus.com). We then looked at the first 5000 results, 
sorted by relevance and selected the 
highest impact articles, according to the following formula: $C / (2016 - Y)^2 \geq a$, where $C$ is the number of 
citations and $Y$ is the year of publication of the article. We set $a=1$ for most cases, with the exception of 
crowdsourcing and social media, where $a=2$, as these research areas are much more populated. The formula was designed 
to oversample recent articles.

Finally, we read the abstracts to manually identify the papers that focused on issues related to 
the functioning and the design of the HMNs. This yielded between 11 and 21 articles (including the seeds) for each HMN 
type.

\begin{table}%
\tbl{Keywords for the Systematic Literature Search\label{tab:keywords}}{%
\begin{tabular}{| p{3.5cm} | p{9cm} |}
\hline
HMN type   & Keywords\\\hline
Public resource computing     & ``public resource comput*'' OR ``volunteer comput*'' OR ``peer-to-peer comput*''\\\hline
Crowdsourcing  & crowdsourcing OR ``crowd sourcing'' OR ``crowd work'' OR ``human computing'' OR ``human computation'' OR ``social 
comput*''\\\hline
Web search engines     & (``search engine'' AND impact) OR (``search engine'' AND design)\\\hline
Crowdsensing    & ``phone sensing'' OR ``mobile sensing'' OR ``crowd sensing'' OR crowdsensing OR ``participatory 
sensing''\\\hline
Online markets   & ``e-commerce'' OR ``prediction market'' OR ``online market'' OR C2C; ``file sharing'' OR 
``peer-to-peer''\\\hline
Social media       & (``social network'' AND site) OR (``social network'' AND online) OR ``social media''\\\hline
Multiplayer online games and virtual worlds     & ``online game'' OR ``multiplayer game'' OR ``virtual world''\\\hline
Mass collaboration & ``open source'' OR wiki OR ``mass collaboration''\\\hline\end{tabular}}
\begin{tabnote}%
\end{tabnote}%
\end{table}%

\end{document}